\documentclass[10pt, conference, compsocconf]{IEEEtran} 
\IEEEoverridecommandlockouts
\usepackage{cite}
\usepackage{amsmath,amssymb,amsfonts}
\usepackage{algorithmic}
\usepackage{graphicx}
\usepackage{textcomp}
\usepackage{xcolor}

\usepackage{enumitem}
\usepackage{subfig}
\usepackage{graphicx}
\usepackage{braket}
\usepackage{physics}
\usepackage{colortbl}
\usepackage{mathptmx} 
\newcommand{\ignore}[1]{}
\usepackage{fancyhdr}
\usepackage[normalem]{ulem}
\usepackage{multirow}

\long\def\comment#1{}

\usepackage{mdwlist}

\def\BibTeX{{\rm B\kern-.05em{\sc i\kern-.025em b}\kern-.08em
    T\kern-.1667em\lower.7ex\hbox{E}\kern-.125emX}}
    
\begin{document}

\title{Understanding Quantum Control Processor Capabilities and Limitations through Circuit Characterization} 

\author{\IEEEauthorblockN{Anastasiia Butko,
George Michelogiannakis,
Samuel Williams, \\
Costin Iancu, 
David Donofrio, 
John Shalf, 
Jonathan Carter and
Irfan Siddiqi\\}
\IEEEauthorblockA{Lawrence Berkeley National Laboratory, 1 Cyclotron Road, Berkeley, CA 94720 \\
Email: \{abutko, mihelog, swwilliams, cciancu, ddonofrio, jshalf, jcarter, iasiddiqi\}@lbl.gov}}

\maketitle

\begin{abstract}
Continuing the scaling of quantum computers hinges on building classical
control hardware pipelines that are scalable, extensible, and provide real time response.
  The instruction set architecture (ISA) of the control processor provides functional abstractions that map high-level 
  semantics of quantum programming languages to low-level pulse generation by hardware. In this paper, we provide a methodology to quantitatively assess the effectiveness of the ISA to encode quantum circuits for intermediate-scale quantum devices with O($10^2$) 
  qubits. The characterization model that we define reflects performance, the ability to meet timing constraint implications, scalability for future quantum chips, and other important considerations making them useful guides for future designs.
  Using our methodology, we propose scalar (QUASAR) and vector (qV) quantum ISAs as extensions and compare them with other ISAs in metrics such as circuit encoding efficiency, the ability to meet real-time gate cycle requirements of quantum chips, and the ability to scale to more qubits.
\end{abstract}

%
\IEEEpeerreviewmaketitle

\maketitle

\section{Introduction}

Inspired by the potential of quantum computing~\cite{Shor,You2011Atomic,Aspuru-Guzik1704} and considering the rising uncertainty of classical computing performance scaling~\cite{7368023}, there has been a rapidly accelerating development of quantum accelerators of different kinds~\cite{Google-2018,IBM-2017,Intel-2018,tqfrigetti,ionQ,dwave}.
Current state-of-the-art quantum computers are typically controlled by an ad-hoc combination of classical control electronics. These classical control electronics are highly diversified and 
see little technology reuse 
across different system implementations~\cite{ionQ, 2013PhRvL.111h0502B}. In all cases, the control electronics must meet stringent real-time and sensitivity constraints that will become increasingly stringent with the growing size of quantum chips. Early experiments all indicate that long-term success of quantum computing hinges on developing better qubit designs combined with a more effective classical control hardware pipeline that is scalable, extensible, and able to provide real-time guarantees~\cite{Quantum_control_survey}.

Although many efforts are underway to build quantum control systems~\cite{intel-chip}\cite{delftqp}, little exists in terms of quantitative performance characteristic and a \emph{regimented method} to tailor digital control logic ISA to best fit the control unit requirements, maximize the efficiency of quantum chips, maximize control unit utilization, and ease qubit scaling. Currently, design criteria for control units focus on metrics that are largely orthogonal to the requirements of large-scale quantum chips with more qubits. Instead, they focus on ease of design or compatibility with existing ISAs to speed up the design cycle, but do so at the cost of sub-optimal control unit performance and scalability.

To address these challenges, our paper takes the first necessary steps towards evaluating the unique characteristics and performance requirements for quantum control units that will dictate the directions of future efforts to optimize their architecture, design and integration.
We have created a systematic approach to the design process and demonstrate the effectiveness of our approach using our own control processor designs. Our implementation extends the open source RISC-V ISA to create a control processor that substantially improves the efficiency of quantum circuit encoding and execution.

\begin{figure*}[htbp]
\centering
\includegraphics[width=\linewidth]{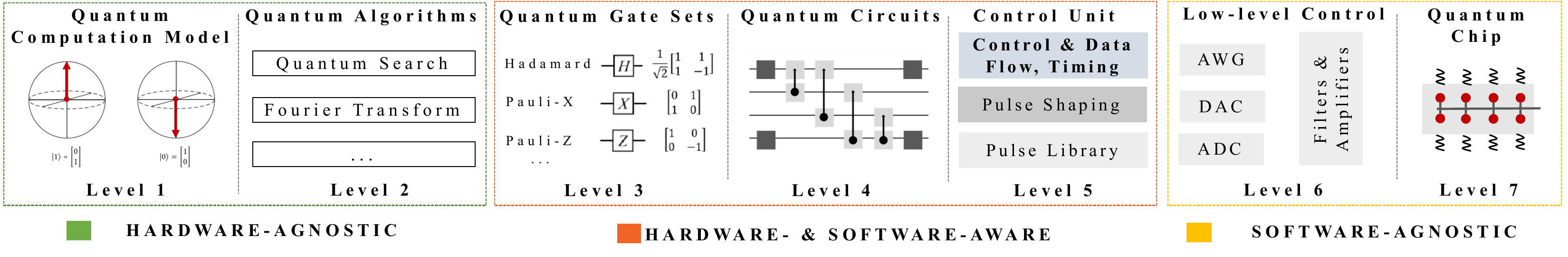}
\caption{Quantum abstraction levels: computation model (Level 1) and algorithms (Level 2) are hardware-agnostic, low-level hardware (Level 6) and quantum devices (Level 7) are software-agnostic. The quantum gate sets (Level 3), quantum circuits (Level 4) and control unit (Level~5) connect software expression of algorithms and the hardware implementation.}
\label{fig:levels}
\end{figure*} 

Namely, our contributions are:
\begin{itemize}[leftmargin=*]
\item We provide a methodology for the exploration of the ISA design space by defining characteristics for operation patterns frequently occurring in quantum algorithms and their circuit structures. The proposed circuit characterization model includes circuit complexity, gate density, gate diversity and gate distribution. We then demonstrate our characterization model by
analyzing two real algorithms, i.e. Quantum Fourier Transform (QFT)~\cite{Shor} and Grover's Algorithm~\cite{1996quant.ph..5043G}. We discuss how our characteristics evolve in Noisy Intermediate Scale Quantum (NISQ) and fault-tolerant quantum reality.

\item We define a set of ISA design requirements that are geared towards quantum real-system needs: support as many qubits, addressing modes, and native gates as possible, while preserving reasonable instruction length, register file size, and pipeline implementation complexity.
\item We propose and implement the QUAntum instruction Set ARchitecture (QUASAR) as a scalar extension to the widely-used open source  RISC-V~\cite{Waterman:EECS-2014-54} ISA, followed by the Quantum Vector (qV) extension to exploit SIMD and MIMD parallelism. Our ISA extensions are publicly available to the community \footnote{https://github.com/lbnlcomputerarch/quasar-spec}.  

\item Using our methodology, we compare four digital control logic ISAs: basic 32-bit RISC-V, eQASM \cite{2018arXiv180802449F}, and the proposed QUASAR and quantum vector (qV) ISA extensions. QUASAR and qV improve code density and execution performance by as much as 15$\times$ and 40$\times$ respectively over baseline 32-bit RISC-V.  
That translates to a 70$\times$ larger encoding capacity compared to alternative ISAs. Finally, we demonstrate QUASAR's ability to guarantee timing constraint satisfaction when running at different control hardware speeds for a set of quantum circuits.    

\item We summarize the capabilities and limitations of existing ISAs based on our motivated metrics, encoding efficiency, generated program size, and potential to satisfy timing constraints for future larger-scale quantum chips. 
\end{itemize}

 The rest of this paper is structured as follows.  Section \ref{sec:levels} provides an overview of the basic concepts of quantum computing theory, related system architecture and software stack. Section \ref{sec:char} characterizes quantum circuits and then motivates our design metrics before presenting our ISA. 
 In section \ref{sec:control}, we discuss the ISA design space and four alternatives for quantum control processor.
 Section \ref{sec:res} shows the results of the ISA analysis, comparative study and real algorithms use cases. This section summarizes the capabilities and limitations of the quantum ISAs. Section \ref{sec:concl} concludes our work and proposes potential future directions.

\section{Background}
\label{sec:levels}
Here, we provide background material for the different quantum abstraction levels shown in Figure \ref{fig:levels}.
It illustrates how hardware-agnostic quantum theory connects to the software-agnostic physical implementation of the quantum chip via intermediate levels. 

\paragraph{Quantum Computation Model}
Quantum bits (\textit{qubits}) are two-level quantum-mechanical systems, whose general
quantum state is represented by a linear combination of two orthonormal basis \textit{states} (basis vectors).  The most common basis is the equivalent
of the 0 and 1 used for bits in classical information theory, respectively $\ket{0} = \begin{bmatrix} 1 \\ 0 \end{bmatrix}$ and
$\ket{1}$ = $\begin{bmatrix} 0 \\ 1 \end{bmatrix}$. The generic qubit state is a superposition of the basis states, i.e. $\ket{\psi} = \alpha \ket{0} + \beta \ket{1}$, with $\alpha$ and $\beta$ complex amplitudes such as $|\alpha|^2+|\beta|^2=1$. The prevalent model of quantum computation is the \textit{circuit model}~\cite{qcircuit}, where information carried by qubits is modified by \textit{quantum gates} that mathematically correspond to unitary operations. A complex square matrix $U$ is unitary if its conjugate transpose $U^*$ is also its inverse, i.e. $UU^* = U^*U = I$.

\begin{figure*}[htbp]
\centering
\includegraphics[width=\linewidth]{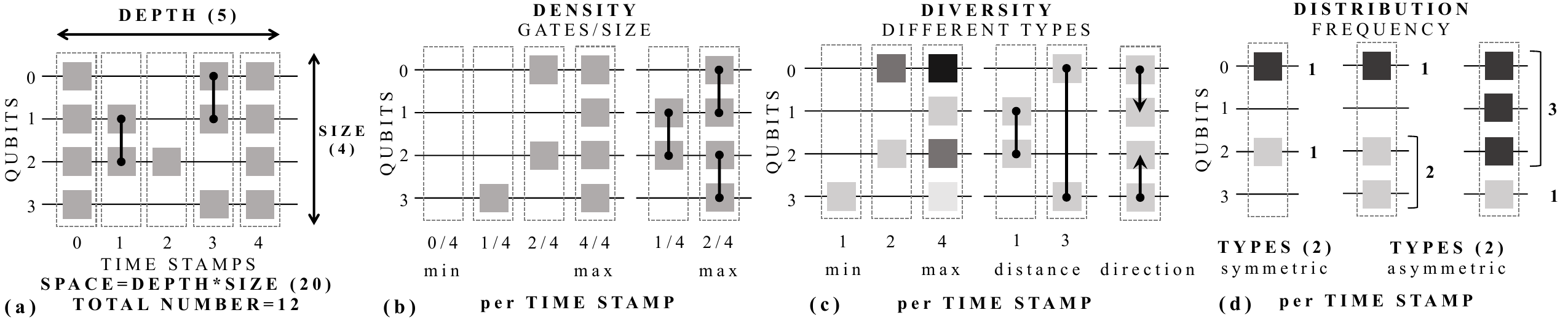}
\caption{Quantum circuit characterization: (a) Circuit complexity in terms of depth and size; (b) Gate density per time stamp; (c) Gate diversity per time stamp; (d) Gate distribution frequency per time stamp.}
\vspace{-0.3cm}
\label{fig:circhar}
\end{figure*} 

\paragraph{Quantum Algorithms}
A quantum algorithm is a step-by-step procedure where in each step a quantum gate can be applied on a quantum bit.
The `Quantum Algorithm Zoo' website cites 420 papers on quantum algorithms~\cite{qzoo}, with popular examples being cryptography, search and optimisation, simulation of quantum systems, and solving large systems of linear equations~\cite{nature-algorithms}. 
 However, the most common algorithms that are often presented as a sub-routine~\cite{Nielsen-2010} are Shor's \textit{quantum Fourier transform (QFT)}~\cite{Shor} and Grover's algorithm for performing \textit{quantum search} \cite{1996quant.ph..5043G}. The Harrow-Hassidim-Lloyd (HHL) algorithm~\cite{hll} usually employs a composition of basic algorithms such as QFT with state preparation. 
 Besides algorithms, practitioners run randomized circuits for hardware characterization using randomized benchmarking~\cite{rb}, quantum volume~\cite{quantumvolume}, or circuits with randomization for error mitigation using randomized compiling~\cite{rc}.
 Quantum supremacy refers to demonstrating that a quantum computer can solve a problem that cannot be practically solved on a classical computer. Google recently claimed quantum supremacy using a superconducting quantum processor~~\cite{qsuprem}.
 In Figure \ref{fig:levels} we depict quantum computational models and algorithms as Level 1 and Level 2 respectively and are \textit{hardware-agnostic}.

\paragraph{Quantum Gate Sets \& Circuits}

A set of quantum gates is {\it universal} if any computation (unitary transformation) can be approximated on any number of qubits to any precision when using only gates from the set. On the high-level programming side, languages provide logical gates that can operate on arbitrary qubits.  However, not all logical gates can be physically implemented on hardware. On the hardware side, a quantum device exposes a minimal set of \textit{native gates} that constitute an universal set.  For example, quantum chips built from superconducting qubits usually provide a gate set consisting of single-qubit rotations (e.g. $R_x(90)$, $R_y(180)$, $R_z(\theta)$) and two-qubit {\it controlled NOT (CNOT)} gates that flip the target qubit {\it if} the control qubit is $\ket{1}$~\cite{Google-2018,IBM-2017,tqfrigetti,qnl}.

While high-level representation of an algorithm with logical gates (quantum circuit) is usually hardware-agnostic, it gradually becomes hardware-aware when going through several stages of transformation to be finally composed of only a set of native gates. In Figure \ref{fig:levels}, these concepts are shown as Levels 3 and 4 and are both \textit{software- and hardware-aware}.

\paragraph{Quantum Control Unit}
The quantum control unit acts as an interface between software and hardware components in the quantum computational stack. This unit manages control and data flow, as well as timing control for algorithm execution. In addition, the unit transforms the circuit into a sequence of control pulses (stored in the pulse library or shaped during execution)
compatible with the low-level control hardware. The quantum execution control unit is shown as Level 5 in Figure \ref{fig:levels} and is \textit{software- and hardware-aware}.  

In order to improve computational capabilities and scalability of a quantum computer, the control unit should be placed in close proximity to the quantum chip. This reduces control pulse propagation delay, improves the ability to control multiple qubits at the same time, and provides fast-feedback for error correction. Recent development efforts~\cite{intel-chip} favor the relevance of this approach. Moving control unit into a low-temperature environment will impose additional design constrains that further emphasize the importance of this work. 

\paragraph{Low-level Control \& Quantum Chip}
Low-level control hardware (backend) is usually composed of Arbitrary Waveform Generators (AWGs), Digital-to-Analog and Analog-to-Digital Converters (DACs and ADCs), as well as numerous filters and amplifiers. 
Quantum chips with superconducting qubits are placed inside a dilution refrigerator and communicate with external components via wires. Low-level hardware and a quantum chip are shown in Figure \ref{fig:levels} as Levels 6 and 7. These components operate at the level of physical pulses with very limited or no notion about computation model or algorithm being run, so they are \textit{software-agnostic}.

The three intermediate levels, i.e. gate set, quantum circuit, and control unit must cooperate in order to provide a suitable interface and advance the state of quantum theory-practice entanglement.
In our work, we propose to evaluate control unit (control processor) capabilities and limitations through quantum circuit characterization.  
\section{Quantum Circuit Characterization}
\label{sec:char}

Quantum circuit characterization allows us to quantify different properties of these circuits that have direct impact on the control unit's ability to handle its main tasks, i.e. control and data flow management and timing control. Moreover, the analysis of the trade-offs in circuit composition can provide useful guidance for compilers and optimizers. 
Unlike other circuit characterization models (coherence, hamiltonian, gate characterization~\cite{Wei:2010:GCF:1837274.1837332}), our proposed model is designed to evaluate the efficiency of quantum circuit encoding.

\subsection{Model Parameters}
\label{sec:parameters}
\textit{Circuit complexity} can be described according to its \textit{size} and \textit{depth}. Figure \ref{fig:circhar} (a) illustrates circuit complexity for an example circuit. The size of the circuit corresponds to the number of qubits it is using, i.e. 4 in this example. The depth of the circuit is the maximum number of gates (grey boxes) that are applied to qubits concurrently at any instance of time (time stamps). The space covered by these two characteristics ($depth \times size$) indicates the maximum capacity per circuit. The total number of gates lies within the boundaries of the circuit capacity.

This common way to characterize circuit complexity does not cover variability introduced by different gate sets and possible gate mappings. Therefore, we introduce additional characteristics used in statistics to quantify these differences. 

\textit{Gate density} indicates to what degree the available gates of the quantum accelerator (capacity) are used by a circuit. In Figure \ref{fig:circhar} (b), gate density is demonstrated at a granularity of \textit{time stamps} (\textit{gates/size}). Gate density varies from low to high in the range from 0.0 (e.g. $\frac{0}{4}$) to 1.0 (e.g. $\frac{4}{4}$). Note, two-qubit gates take twice as much space and fill the capacity `faster'. 

\textit{Gate diversity} represents the number of different (unique) gates in a circuit or per time stamp. Figure \ref{fig:circhar} (c) shows per time stamp gate diversity. The lowest level of gate diversity refers to a single type. The highest level of gate diversity implies that at time stamp \textit{ts} the number of gate types equals the total number of gates. 

These minimum and maximum boundaries are theoretical and can be different in practice with native gate sets and circuit sizes. A typical superconducting native gate set can have around seven constant single-qubit operations (i.e. $R_x$ and $R_y$ rotations, measurement). The arbitrary $R_z(\theta)$ rotation can create a large number of unique operations defined at the user-level thereby increasing possible diversity even for a small native gate set. With one type of two-qubit gates, the differences in circuit structure can be expressed by \textit{distance} between target and control qubits or by \textit{direction}. These differences strongly depend on real chip connectivity.

\textit{Gate distribution} shows the count of the occurrences of each gate type in a circuit of within a time stamp. In statistics, this metric is known as frequency distribution. A discrete uniform distribution or \textit{symmetric distribution} has an equal number of gates of each type. In a \textit{asymmetric distribution}, the number of gate occur at irregular frequencies per type, and \textit{skewness} is a measure of the asymmetry of the distribution. Figure \ref{fig:circhar} (d) shows both symmetric and asymmetric gate distributions with different levels of asymmetry. 

\subsection{Real Algorithms and their Circuits}
\label{subsec:real-al}

Our proposed metrics model characterizes a variety of quantum circuits. Here, we discuss how these characteristics relate to the reality of quantum algorithms when applied to NISQ devices, as well as projecting them into future error-corrected quantum devices possibly of larger scales.

NISQ devices have limitations. Namely, noise and short qubit lifetime impose restrictions on the circuit structure to be implemented on the device.  
Superconducting qubits have inconveniently short lifetimes~\cite{2019arXiv190513641K}. To maximize the use of available qubits (capacity), quantum circuit density  
needs to be as high as possible. This minimizes idling qubits. On the other hand, parasitic crosstalk and other environmental noise require gate pulses to be separated in time and space~\cite{2019PhRvP12e4023M}. This motivates a circuit structure implementation that is significantly more sparse. Finding the trade-off between strong qubits interaction and low crosstalk prevails in NISQ experiments.

Beyond NISQ, error correction is essential in achieving fault-tolerant quantum computation. Classical error correcting codes use so-called \textit{syndrome} measurement to diagnose an error (error detection) and then reverse it by applying a corrective operation based on the syndrome (error correction). These measurement-correction operations need to be applied constantly and are typically independent from the algorithm. If this approach is applied to quantum, the resulting circuits become much more dense and diverse~\cite{2009arXiv0904.2557G}.   

\begin{figure}[t]
\centering
\includegraphics[width=\linewidth]{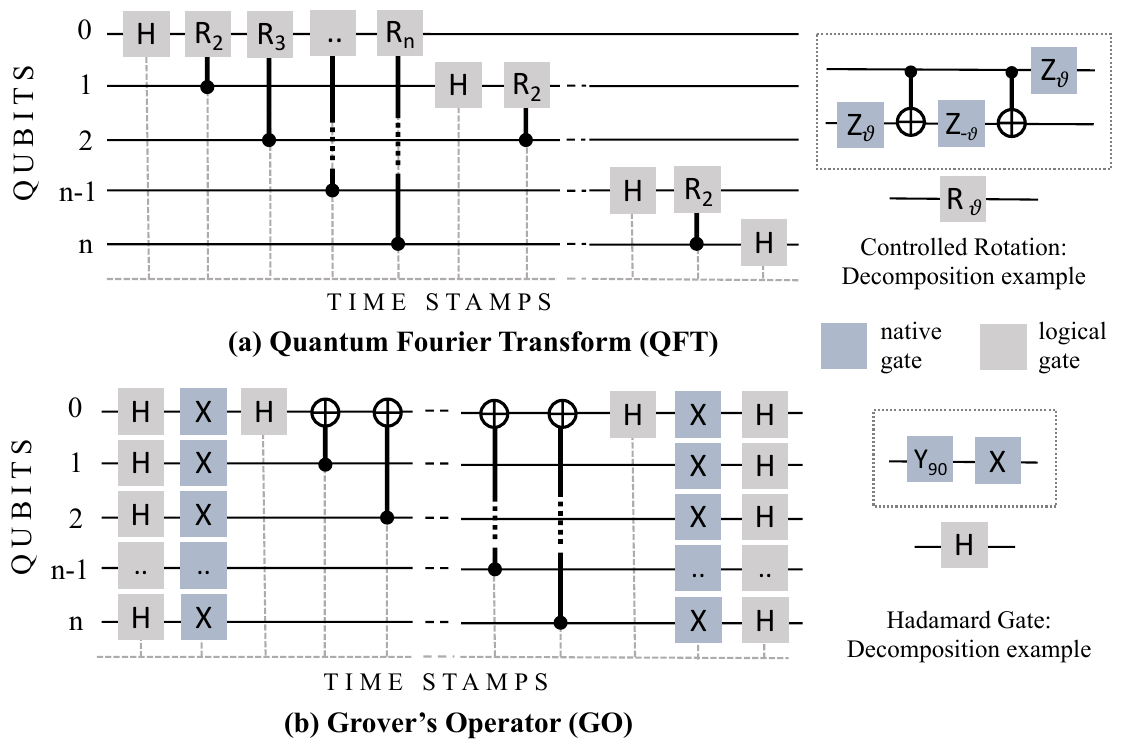}
\caption{Algorithms circuits: (a) Quantum Fourier Transform (QFT) and (b) Grover's Operator (GO).}
\label{fig:al-cir}
\end{figure}

Figure \ref{fig:al-cir} shows two example circuits: QFT and Grover's Operator (GO) that is a sub-circuit of Grover's algorithm. These two algorithms are used in a large variety of problems. QFT is the quantum analogue of the inverse discrete Fourier transform. It is a key building block for many existing quantum algorithms, such as Shor's algorithm and quantum phase estimation \cite{Nielsen:2011:QCQ:1972505}. The QFT circuit is composed of two quantum logic gates: the Hadamard gate (\texttt{H}) and the Controlled Phase gate (\texttt{CF}). These two types of gates are decomposed into a subset of base rotations. Grover's algorithm \cite{1996quant.ph..5043G} is a quantum search algorithm that solves the problem in the order of $O(\sqrt{N})$ versus $O(N)$ operations on a classical computer. The algorithm is composed of repeated applications of a quantum subroutine called \textit{Grover operator} \cite{Nielsen:2011:QCQ:1972505}. The Grover operator is built out of Hadamard gates surrounding an operation that performs a conditional phase shift. Similar to the QFT, the Hadamard and phase shift gates are decomposed into simple rotations and controlled-not gates.  

\begin{figure}[t]
\centering
\includegraphics[width=\linewidth]{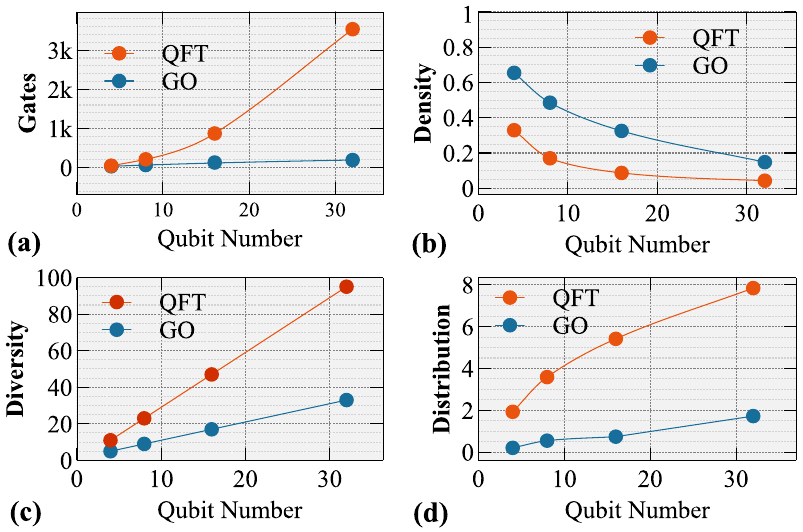}
\caption{Circuit characterization: (a) total gate number, (b) gate density, (c) diversity, and (d) distribution asymmetry for QFT and GO.}
\label{fig:al-char}
\end{figure}

Figure \ref{fig:al-char} compares QFT and GO circuit characteristics when varying  the number of qubits. QFT shows a significantly larger increase in the number of gates than GO. At the same time, the QFT circuit has lower density. Diversity is composed of arbitrary Z rotations and CNOT gates of different distances for QFT, while GO differs in CNOT distance only. With more qubits, the number of CNOTs in QFT's circuit begins to prevail. In turn, this causes a significant increase in distribution asymmetry. GO shows a similar trend but at a smaller scale. These characteristics when applied on real hardware will vary depending on implementation capabilities.

Eliminating noise will further reduce density (Figure \ref{fig:al-char} (b)) because environmental noise requires gate pulses to be separated in time and space. However, circuit depth cannot stretch beyond qubit lifetime. At that time, density will become zero. A chip's physical qubit connectivity affects all four characteristics because it requires moving a qubit's state (due to the no-cloning theorem \cite{Wootters1982Single}) to enable two-qubit interaction since these two qubits need to be physically connected. Qubit state movement is done by a SWAP operation that can be decomposed in multiple gates. That significantly increases the number of applied gates and stretches CNOT operation over multiple time stamps resulting in increased sparsity, lower diversity, and higher distribution asymmetry. The native gate set determines gate decomposition as shown in the example in Figure \ref{fig:al-cir} (Hadamard gate (\textbf{$H$}) and controlled rotation (\textbf{$R_n$})). In turn, that affects diversity as well as circuit depth. If a Hadamard gate can be implemented as a single operation (pulse), it requires fewer time stamps to be executed, which means lower depth and higher density.

\section{Control Processor Design}
\label{sec:control}

A quantum system composed of a control unit and a quantum chip is a \textit{real-time system}~\cite{10.1093.comjnl.9.3.249}.
In the context of quantum computing, a control unit sends control pulses, receives measurement data, processes them, and provides feedback in exact correspondence to the algorithm circuit structure. Control and feedback responses should be guaranteed within a specific timing constraint. Because qubits have finite lifetimes, the usefulness of the computation result decreases abruptly and may become negative (false results that can lead to miss-interpretation) if tardiness increases. The \textit{timing constraint} is defined by the period of time in between two time stamps. The lower bound of the timing constraint is equal to the duration of the shortest gate pulse (e.g. single-qubit operation in superconducting qubits). 

\begin{figure*}[t]
\centering
\includegraphics[width=\linewidth]{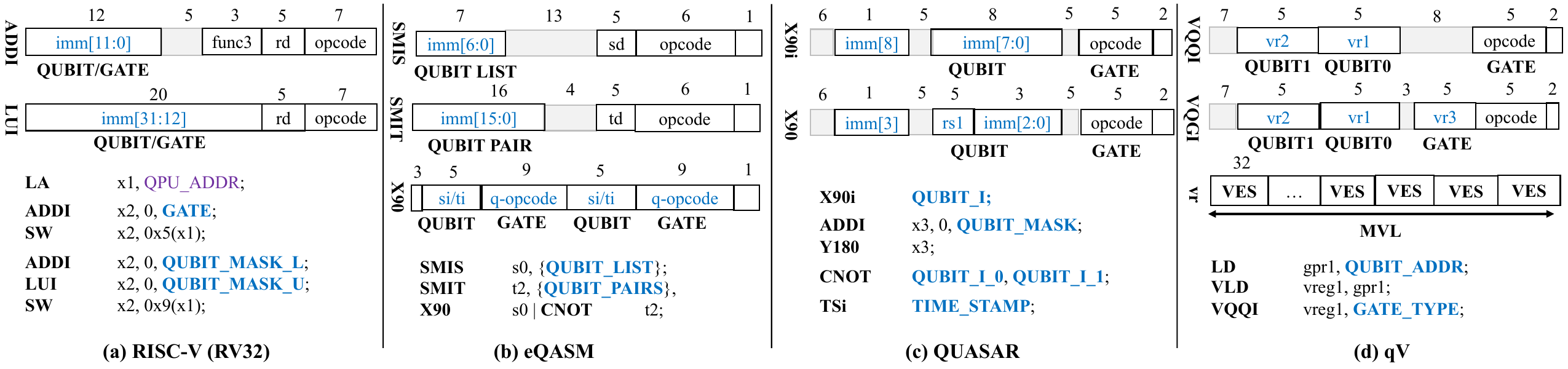}
\caption{ISA encoding for quantum operations with assembly code examples: (a) base (not-extended) 32-bit RISC-V (RV32) via memory-mapped IO communication (MMIO), (b) eQASM, (c) our QUASAR extension to RV32, and (d) our double-indexed vector extension qV.}
\vspace{-0.3cm}
\label{fig:isa-des}
\end{figure*}

\textit{A quantum control processor} is a hardware implementation of the control unit shown in Figure \ref{fig:levels} (e.g. control \& data flow, timing). It plays a key role in orchestrating timely gate application and maintaining efficient control and data flow during circuit execution. The Instruction Set Architecture (ISA) is a core component of a processor. Existing ISAs take into account energy constraints for mobile compute or floating-point operation efficiency for high-performance servers~\cite{6522302}. Both of these design constraints are irrelevant in the context of quantum control.

\subsection{ISA Alternatives}
\label{subsec:isa-alt}

Operation patterns in quantum circuits can be found at vertical sub-circuits that correspond to a single time stamp. Based on Flynn's taxonomy~\cite{5009071}, such a sub-circuit expresses different forms of parallelism, i.e. SIMD (single gate, multiple qubits) or MIMD (multiple gates, multiple qubits). The encoding must balance multiple competing requirements: (i) support as many qubits, addressing modes, and native gates as possible, (ii) evaluate the impact of register size, qubit addressing, gate type fields on the instruction size, and (iii) try to have instructions encoded into a format that will be easy to handle in a pipelined implementation. We study four architecture alternatives (two existing approaches and two proposed approaches) that balance these requirements in different ways.

\textit{RISC-V RV32} is a RISC ISA~\cite{Waterman:EECS-2014-54,rv-cores} that does not explicitly support quantum operations.
We use RISC-V as \textit{a baseline} to
motivate specialized quantum-oriented extensions.

\textit{eQASM}~\cite{2018arXiv180802449F} is a 32-bit ISA that targets a seven-qubit superconducting quantum processor with a two-dimensional square lattice connectivity~\cite{2017PhRvP8c4021V}. eQASM consists of four major groups of instructions: control flow, data flow, arithmetical and logical operations, and quantum-related instructions. Quantum-related instructions include quantum \texttt{WAIT} of immediate and register formats to control timings, instructions that specify a target qubit for gate application (\texttt{SMIS} for single-qubit ID list and \texttt{SMIT} for two-qubit pair list,) and gate application instructions that can fit two types of gates (two gates, multiple qubits).

\paragraph{Proposed Approaches:}
We propose two ISAs that differ from existing approaches in their quantum operation encoding efficiency, in their support of larger number of qubits, and in being topology-agnostic. 

\textit{QUASAR} is a SIMD quantum extension to the RISC-V ISA that explicitly supports quantum operations (gates) and timing control. In contrast to eQASM, we prioritize the ability to control a large number of qubits in a scalar format (single gate, multiple qubit) with different addressing modes and provide a generic (topology-agnostic) mode to address two qubits in a two-qubit gate. Gate instruction scheduling and decomposition for a specific connectivity is done by the compiler. With a fixed 32-bit instruction length, QUASAR supports up to 512 qubits (compared to seven for eQASM), two addressing modes (immediate and mask), and up to 15 unique gates (single-qubit, two-qubit, arbitrary rotation, measurement).

\textit{qV} is vector version of QUASAR that uses a double index format that leverages and extends vector instructions to address both forms of parallelism (SIMD and MIMD), prioritizing control throughput over processor design complexity.

qV adds two new instruction classes for quantum gates. The first ({\tt vqqi}) is a two-register format with an {\tt op} field to specify a common gate.  Such an instruction takes the two lists of qubits (two vectors of indices) and applies the same operator (gate) to all of them. In essence, this expresses the traditional SIMD-style approach to parallelism.
The second form ({\tt vqqg}) is a three-register format. As before, we have two lists of qubits (vector registers encode indices of qubits) but augment this with a third register that encodes a list of gates to be applied ({\tt gate[vreg3[]] qubits[vreg1[i]], qubits[vreg2[i]]}). This novel form combines two qubit index lists with a quantum gate list easily expressing MIMD parallelism (any combination of qubits and gates).  Masking can be effected with either a {\tt nop} field in the third register or enumerating qubits from 1 (qubit$_1$ is the first qubit) and treating any index of zero in the first register as a {\tt nop}.

\subsection{ISA Encoding Analysis}
\label{subsec:isa-an}

Depending on the ISA, the number of instructions per sub-circuit will significantly vary. Both the ISA and processor implementation (pipeline stages, memory hierarchy, register file size, etc.) directly affect whether or not the control processor will meet timing in order to deliver the gates for the next time stamp cycle. Figure \ref{fig:isa-des} shows encoding formats and assembly code examples for all four design alternatives. Encoding fields and assembly instructions colored in blue relate to quantum operations. Encoding fields colored in grey show unused instruction space.

\textit{RISC-V RV32} ISA (a) does not support quantum operations, thus it needs to control an external device (qubits) using classical mechanisms, such as hardware interrupts, port-mapped I/O (PMIO) that uses special instructions, or memory-mapped I/O (MMIO) that assigns memory addresses to I/O devices. A hardware interrupt is device-initiated and unidirectional; it can be used only to indicate that the quantum device requires processor attention. I/O operations have historically caused performance concerns and can also slow memory accesses if memory and I/O operations share common buses. Here, we focus on MMIO as the most suitable for quantum control. Using Direct Memory Access (DMA) instead would follow a similar access pattern.

MMIO requires the processor to perform the following sequence of instructions: load a quantum processing unit address into a register (x1), load a control value into a register (x2), store in x2 to the address  in x1. Depending on the control value range (a qubit number or a gate type), loading can be done by a \texttt{ADDI} instruction that contains a 12-bit immediate field or as a combination of \texttt{ADDI} and \texttt{LUI} that loads the upper 20 bits with an immediate field to fill the most significant bits of the 32-bit register. Thus, the qubit address space and gate type space varies from $2^{12}$ to $2^{20}$ if using direct addressing or from $12$ to $32$ if using mask addressing. A single-qubit operation with timing control will require 8-9 instructions and at least 2 registers, and a two-qubit operation 10-12 instructions and 2 registers.

\textit{eQASM} (b) performs the following sequence of commands: it loads a \textit{qubit list} with a \texttt{SMIS} instruction or a \textit{qubit pair} with a \texttt{SMIT} instruction into a register, and then it executes a quantum operation on the value from this register. A quantum operation instruction can fit up to two gate/register pairs. Thus, eQASM's qubit address space ranges to up to 7 qubits with only the mask addressing mode available; the gate type ranges to up to $2^9$. eQASM supports 16 two-qubit interactions as the chip connectivity is encoded in a 16-bit immediate value (QUBIT PAIR). Both single-qubit and two-qubit operations with timing control require 3 instructions and 2 registers.

Requiring the ISA to be topology-aware dramatically limits its flexibility. First, it requires continuous changes to the instruction format, processor micro-architecture, and compilers every time the topology or size of the quantum chip changes, even if the change is minimal.
Second, if we generalize and require the ISA to be able to express all possible couplings, not only those physically present, we end up dealing with the complete graph. The number of edges of a complete graph with $n$ vertices is calculated as $n*(n-1)/2$. Taking into account that each edge has two directions, this number is multiplied by $2$. For a reference 7-qubit chip, the number of bits required to cover any possible topology becomes 42 vs. 16 and already does not fit into the 32-bit instruction format. For a 32-qubit chip, the number goes up to 992 bits.

\textit{QUASAR} (c) supports two qubit addressing modes, i.e. direct (immediate) and mask format. Both gate type and the number of qubits are encoded within a single instruction. In direct addressing mode, a qubit ID (index) is held in a 9-bit immediate field. In mask mode, multiple qubit IDs are pre-loaded into a register as a 32-bit mask with a 4-bit immediate value to indicate the offset (\textit{sliding mask} approach). The gate type is encoded in a 5-bit opcode field. Thus, the qubit address space ranges to up to 512 qubits; the gate type number ranges to up to $2^5$ with two instructions for reserved for timing control. A single-qubit operation with timing control requires 2-3 instructions with 0 or 1 registers in use. A two-qubit operation requires 2-4 instructions with 0-2 registers in use.

\begin{figure*}[t]
\centering
\includegraphics[width = \textwidth]{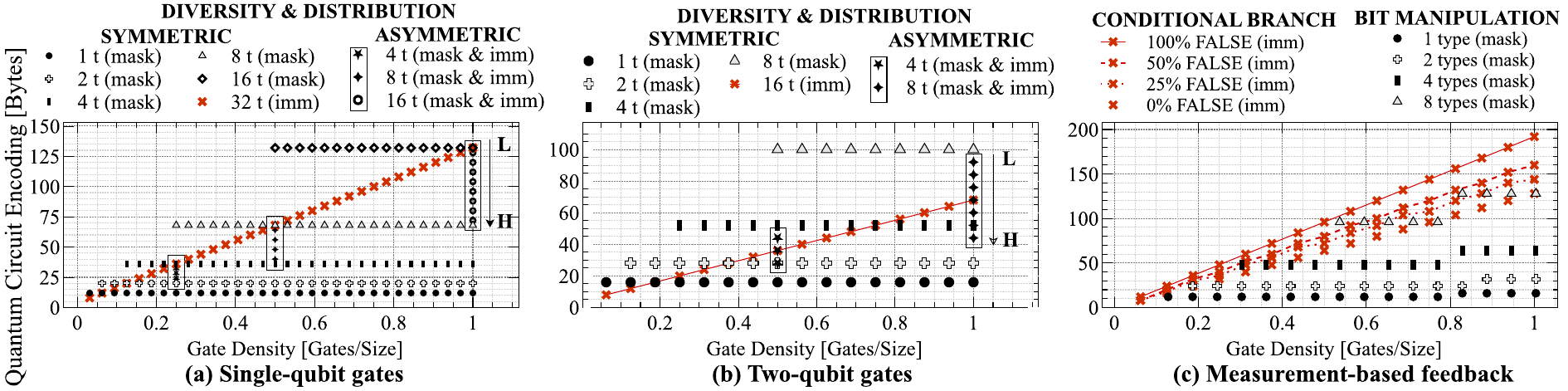}
\caption{Quantum circuit encoding with QUASAR ISA: synthetic circuits with varying density, diversity and level of asymmetry.}
\vspace{-0.3cm}
\label{fig:quasar}
\end{figure*}

\textit{qV} execution of a quantum circuit will nominally iterate on three ({\tt vld, vld, vqqi}) or four ({\tt vld, vld, vld, vqqg}) instructions depending on whether the instruction stream is attempting to exploit SIMD or MIMD parallelism (one must load the vectors of indices of the qubits, the vector of gates, and execute vector instruction). An example is shown in Figure~\ref{fig:isa-des} (d). The qubit address space and gate type number depend on the Vector Element Size (VES) that is equal to $2^{VES}-1$ with 0 reserved for a \texttt{NOP} operation. The Maximum Vector Length (MVL) indicates the maximum value of VES in a vector register file. A single-qubit operation requires 4 instructions and 2 vector registers. A two-qubit instruction requires 6 instructions and 4 vector registers.

We note that vector architectures have a larger set of parameters. Therefore, one should tailor MVL to match the typical SIMD or MIMD parallelism available in a quantum circuit (maximum number of gates in a window that address different qubits), as well as VES used for qubit indices to support varying numbers of qubits.

\section{Evaluation Results}
\label{sec:res}

\subsection{Experimental Setup}

\underline{Evaluation Metrics:} Our study focuses on two performance metrics: \textit{encoding efficiency} and \textit{execution time}. Encoding efficiency represents how many \texttt{bytes} are required to represent a circuit. Higher encoding efficiency (fewer bytes) means a lower instruction count and fewer data to be moved from memory to registers.

We use execution time to evaluate satisfying timing constraints. To satisfy timing constraints in every time stamp, the execution time of the related sub-circuit has to be less then a threshold value. 
We calculate execution time using in part the Instructions Per Cycle (IPC) rate produced by an in-order 5-stage pipelined processor.

\underline{Benchmarks:} Our experiments include two types of circuits: \textit{synthetic circuits} and circuits based on \textit{real algorithms}. Synthetic circuits are built to explore the range of the metrics we define in Section~\ref{sec:parameters}. 
Our circuits are composed of single- and two-qubit gates representing the variety of possible combinations in a three-dimensional space created by our three metrics.
Real algorithms include \textit{Quantum Fourier Transform (QFT)}~\cite{Nielsen:2011:QCQ:1972505}  and \textit{Grover's operator} \cite{Nielsen:2011:QCQ:1972505} shown in Section \ref{subsec:real-al}.

\subsection{Encoding Efficiency}

\underline{Single- and Two-qubit Gates:}
Figures~\ref{fig:quasar} (a) and (b) show QUASAR single- and two-qubit gate sub-circuit encoding. Encoding results include two qubit addressing modes (immediate and mask) applied on sub-circuits with varying gate diversity (from $\frac{1}{32}$ to $\frac{32}{32}$) with symmetric distribution and asymmetric distribution. 
The red-cross line (i.e., 32 types) represents the worst case scenario in terms of gate diversity encoded with the immediate instruction format. The encoding shows linear dependency on \textit{gate density} such as 1 instruction or 4 Bytes per gate plus 1 instruction or 4 Bytes to increment the time stamp. The code size goes up to 132 Bytes for a 32 single-qubit sub-circuit  
and up to 68 Bytes for 16 two-qubit sub-circuit. All gate diversity scenarios with a \textit{symmetric distribution} are encoded using the mask register format. Since all qubits are located within the 32-bit range, gate density does not impact encoding efficiency.  
The code size ranges from 12 Bytes to 132 Bytes for single-qubit gate sub-circuits and from 16 Bytes to 100 Bytes for two-qubit gate sub-circuits. In scenarios above the red line, mask encoding is less efficient than immediate format and vice versa. This trend is significantly larger for two-qubit gates.
Circuits with gate diversity of half of the possible maximum, i.e. 8 types and symmetric distribution, do not benefit from using a mask even at the largest gate density of 16 gates. Also, only half of the asymmetric circuits can be efficiently encoded with the combination of two formats. The reason lies in the additional mask load. The demonstrated model assumes that a 32-bit mask can be loaded with a single \texttt{la} instruction. This load requires 4 bytes of encoding space, but can cause longer execution due to memory latency. An alternative load is composed of two immediate instructions \texttt{lui} and \texttt{addi} that do not require memory access, but take 8 bytes to encode the sequence.

In the case of asymmetric distributions, a combination of two addressing modes results in a significant reduction in the requisite number of instructions. The reduction increases with increasing level of asymmetry. For example, if there are 8 gate types per 16 gates (0.5 gate density), maximum reduction is achieved when the circuit fragment is composed of 7 gates of 7 different types encoded with the immediate format and 9 gates of the same type encoded with the mask format.

As described in Section \ref{subsec:isa-an}, qV vector extension requires 4 instructions for single-qubit and 6 instructions for two-qubit gates. The vector format requires a general purpose register to hold the memory address of each vector register used. qV enables  SIMD  and  MIMD  (different  gates  on  different qubits) parallelism in a single instruction. Moreover, it also allows to encode single- and two-qubit gates in a single instruction. Encoding efficiency is independent of circuit characteristics, thus the code size for any configuration is 16 Bytes for a single-qubit gate sub-circuit and 24 Bytes for a two-qubit gate sub-circuit. However, each operation requires massive data movement. A single data vector block of 32 MVL and 32 VES takes 128 Bytes. Thus a \texttt{vqqi} instruction requires 2 vector registers or 256 Bytes and a \texttt{vqqg} instruction requires 3 vector registers or 384 Bytes. 

Within the design goal, i.e. 7 qubits and 16 couplings, eQASM provides good scalability with the code size ranging from 12 Bytes for 1-type  (low diversity) single-qubit gate sub-circuits and for multi-type (high diversity) two-qubit gate sub-circuits, to 48 Bytes for multi-type (high diversity) single-qubit gate sub-circuits. To support a larger number of qubits with eQASM, the control system will need an additional sequencer to shift the qubit address range or multiple processing units running in parallel.

Finally, based on the analysis from Section \ref{subsec:isa-an}, \textit{RISCV-V RV32} requires 8-9 instructions for a single-qubit gate and 10-12 instructions per a two-qubit gates. Every additional single-qubit gate type requires at least 4 instructions. Code size ranges from 36 Bytes to 656 Bytes for single-qubit gate sub-circuits with increasing diversity and from 40 Bytes to 1040 Bytes for two-qubit gate sub-circuits.

\underline{Measurement-dependent feedback:}
It refers to circuits with conditional gate applications like in \textit{Quantum Teleportation}~\cite{10.5555.1972505} or error correction code~\cite{Gottesman:2014:FQC:2685179.2685184}. While this pattern is less relevant for NISQ devices, it is crucial for fault-tolerant quantum computations. 

Figure~\ref{fig:quasar} (c) shows the encoding of quantum circuit fragments composed of two time steps: qubit data measurement and measurement-dependent feedback. We evaluate two encoding approaches. The first approach uses QUASAR RV32 conditional branch instructions and the immediate single-qubit addressing mode. Results are illustrated with red-cross lines. The application of measurement gates is similar to single-qubit gate encoding. Thus, for the measurement-dependent feedback model, we use the circuit fragment that contains only these instructions related to conditional branches and feedback gates. Different red-cross lines indicate the percentage of conditions to be \texttt{TRUE} or \texttt{FALSE}. If the condition is \texttt{FALSE}, the feedback gate is applied resulting in additional instructions execution. Otherwise, the only instruction to be executed is the conditional branch.

An alternative approach uses the mask addressing mode and bit manipulation techniques on circuits with a gate diversity of 1 type, 2 types, 4 types and 8 types. Here, conditional branch instructions are eliminated. In case a pattern of 1, 2 or 4 types is present in the measurement-dependent feedback circuit, such a circuit can be encoded up to 12$\times$, 6$\times$ and 3$\times$ times more efficiently respectively. This reduction can be achieved by using only standard logical operations, such as \texttt{and}, \texttt{or}, \texttt{not}, left and right shifts, etc. Using an explicit bit manipulation extension such as Bitmanip \cite{bitmanip}, measurement-dependent feedback can be encoded even more efficiently. At the end, the feedback quantum instruction is executed despite the measured data. However, the mask register indicates whether or not the gate has to be executed.
Code size ranges from 8 Bytes to 192 byte.

In eQASM ISA, there are three factors that make eQASM feedback control encoding with branches less efficient than quantum RISC-V based ISAs (QUASAR or qV). First, the processor does not have direct access to quantum data. Initially, measured data first arrive to the special-purpose register, then an additional move-like instruction has to be executed to copy data to a general purpose register. After that, the data can be analyzed for conditional branch execution. As a second factor, eQASM operates on a 7-qubit/bit granularity, but it is a 32-bit instruction set, thus all classical instructions, such as logical and arithmetical operations are of 32-bit granularity. Moving and operating on a 7-bit value within the 32-bit register is unproductive. The third and last factor is that eQASM lacks several crucial logical operations, such as left and right shifts to enable efficient bit manipulation.

\begin{figure}[t]
\centering
\includegraphics[width=\linewidth]{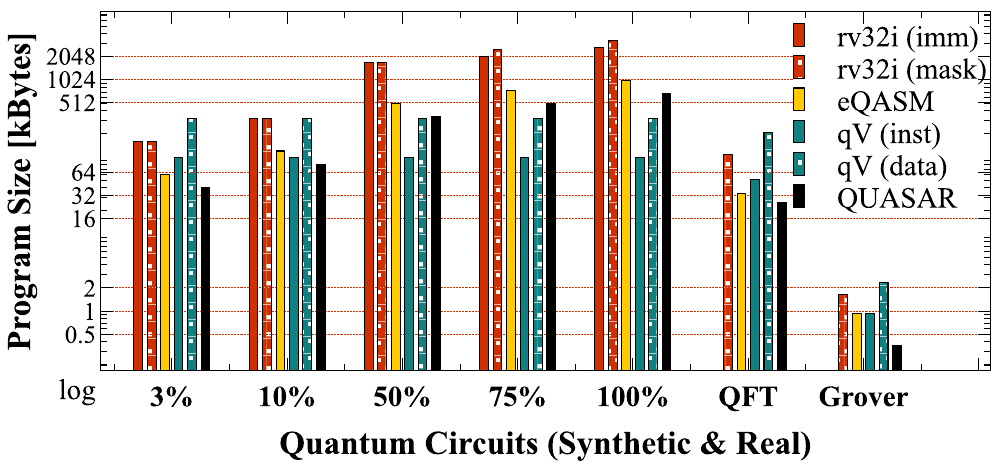}
\caption{Program size of the shown quantum circuits that have varying density and high gate diversity.}
\label{fig:program-size}
\end{figure}

\subsection{Program Size}
We examine program size estimates for two groups of quantum circuits. The first group consists of synthetic circuits with different levels of single-qubit gate density, i.e. 3\%, 10\%, 50\%, 75\% and 100\%. Circuit depth is calculated assuming that the qubit lifetime is 100us and the gate time is 20ns. The second group of circuits consists of real algorithm sub-routines, i.e. QFT and Grover's operator shown in Figure \ref{fig:al-cir}. In both cases, the circuit size is 32 qubits. 

Figure \ref{fig:program-size} shows the program size for each circuit encoded for each ISA.
We highlight three categories in the logarithmic scale that correspond to the following memory requirements: 
low (below 2 kB), 
medium (16-32 kB corresponding to a L1 cache), 
and high (512-2048KB corresponding to a L2 cache).
Note that qV data is split into two bars corresponding to instruction and data. 

With 8 instructions per single-qubit operation,  RV32I has the highest memory requirements among all ISAs. Even with a gate density as low as 3\%, the program will rarely fit into a typical L1 cache. 
Moreover, with a gate density of 50\% or higher, the program will not even fit into the typical L2 cache. 
The non-deterministic behaviour of caches and high miss penalties 
increase the risk of timing failures.
By contrast, eQASM can encode low density circuits within L1 cache capacities. However, because of its restriction to 7-qubit quantum devices, program size grows rapidly with increasing gate density. 

qV encoding is independent of circuit characteristics. Program instruction sizes are comparable to a typical L1 cache size, and data requirements are below 512MB. qV provides the best encoding for high density circuits and its memory requirements do not grow with the number of qubits, making it the natural choice for future larger quantum accelerators.
In contrast, QUASAR provides the smallest program size for most circuits, but suffers compared to qV for high density circuits (starting from 50\%).

The real algorithm circuits have relatively low gate density (see Section \ref{subsec:real-al}). Here, program size is comparable to a typical L1 cache size. However, these circuits represent only a fragment of the algorithm, thus program size will grow proportionally to application length. Moreover, for real systems, we expect gate density to be higher due to the need for additional gates to mitigate errors and stabilize the circuit. This further motivates qV.

\begin{figure}[t]
\centering
\includegraphics[width=\linewidth]{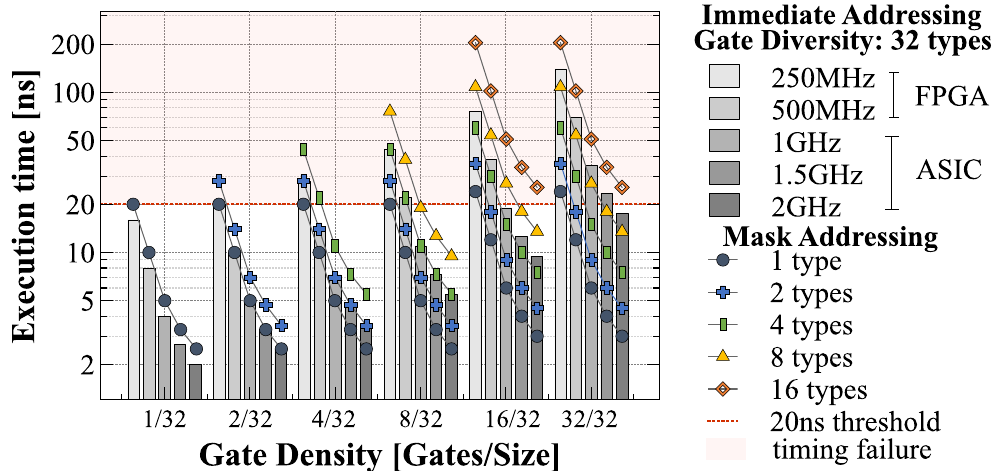}
\caption{Timing of a single quantum circuit. Execution time is estimated based on processor frequencies for FPGAs and ASICs. A 20ns threshold corresponds to the duration of a single-qubit gate. Cases below the threshold line indicate processor failure to deliver control signals on time.
}
\label{fig:cs4}
\end{figure}

\subsection{Timing Constraint Satisfaction}
\label{subsec:tcs}
 
Given the number of cycles for each circuit scenario, i.e. immediate addressing mode for the highest diversity circuits of 32 types and mask addressing mode for 1-, 2-, 4-, 8- and 16-type diversity circuits, we report the time per single time stamp execution depending on processor speed.
We consider 200 MHz and 500 MHz that represent the execution on an FPGA board and 1 GHz, 1.5 GHz and 2 GHz for a typical ASIC implementation.
Grey bars represent execution time for immediate addressing mode and marked lines represent execution time for mask addressing mode. The red line at 20ns corresponds to the duration of a typical single-qubit gate in superconducting technology. The area above the threshold line indicates timing constraint satisfaction failure. In all cases located in this area, the processor fails to deliver control gates on time resulting in circuit execution inaccuracy and erroneous results. 

As shown, the processor running on an FPGA can guarantee timely gate delivery for up to a density of eight gates with the immediate format and up to a 32-gate density with the mask format if there are only 2 types of gates. ASIC implementations running at around 2GHz allow executing any 32-qubit circuit using the immediate format or mask for circuits with a gate diversity of 8 types and lower. 

This illustration allows us to indicate when the processor will fail depending on algorithm characteristics. According to our circuit characterization in Section \ref{subsec:real-al}, QUASAR can guarantee timely execution of QFT, GO, or any other circuit that lies within the boundaries of upper mentioned gate density and diversity ranges.

\subsection{Summary}
In this section, we summarize the trade-offs of each ISA.

\textbf{RV32:}\\
    \underline{pros}: Does not require any changes in software or hardware. Supports various extensions that provide benefits for quantum computing, i.e. bit manipulation and floating-point.\\
    \underline{cons}: Inefficient quantum circuit encoding requires a large number of instructions per gate that (1) demand high processor instruction throughput, and (2) result in program sizes larger than typical instruction cache capacities. Meeting timing constraints can only be guaranteed for simple circuits.

\textbf{eQASM:}\\
\underline{pros}: Provides efficient encoding for a 7-qubit chip with a two-dimensional square lattice connectivity. 
Is flexible in terms of quantum gate operations allowing one to redefine the set of gates at compile-time. Supports reduced MIMD paralellims (two gates, multiple qubits).\\
\underline{cons}: Encoding is narrowly focused and does not easily adapt to quantum systems with more qubits or different connectivity. 
Masking of each bi-directional connection is expensive and does not scale. Does not support bit manipulation.
Requires a specialized software ecosystem.

\textbf{QUASAR:}\\
\underline{pros}: Encoding supports both qubit addressing modes (immediate and mask) which in turn allows QUASAR to support a wide range of quantum circuits. Supports bit manipulation. Easily integrates with the existing RISC-V ecosystem.\\
\underline{cons}: Number of qubits is fixed. Does not support MIMD parallelism (multiple gates, multiple qubits).

\textbf{qV:}\\
\underline{pros}:
Enables SIMD and MIMD (multiple gates, multiple qubits) parallelism in a single instruction. 
Encodes single- and two-qubit gates in a single instruction.
Encoding efficiency is independent of circuit characteristics.
Can easily and inexpensively be scaled to support an arbitrary number of qubits.\\
\underline{cons}:
Requires more data movement and more memory capacity per gate regardless of circuit.
Vector element size (VES) is $log_2$ of the maximum of the number of qubits  and the number of unique gates.
Vector register-based ISA adds substantial complexity and hardware cost.
Requires a vector extension to the RISC-V software ecosystem.
\section{Conclusions}
\label{sec:concl}

In this paper, we develop an evaluation methodology and circuit characterization model for quantum ISAs that defines four key quantum circuit characteristics: circuit complexity, gate density, gate diversity, and gate distribution frequency. We define quantum ISA design requirements dictated by the timing constrain of the real-time nature of quantum system. Based on these, we propose and demonstrate the scalar QUASAR and the vector qV ISA extensions to RISC-V. 

\section*{Acknowledgement}
The research leading to these results has received funding from the the U.S. Department of Energy, grant agreement n\textsuperscript{o} DE-AC02-05CH11231. 
  
\pagebreak


{\footnotesize
\bibliographystyle{IEEEtran}
\bibliography{main,quantum} 
}

\end{document}